\documentclass{ptapap}
\author{A. Pigulski}[IAUWr]
\author{M. Jerzykiewicz}[IAUWr]
\author{M. Ratajczak}[IAUWr]
\author{G. Michalska}[IAUWr]
\author{E. Zahajkiewicz}[IAUWr]
\author{the BRITE Team}
\affil[IAUWr]{Instytut Astronomiczny, Uniwersytet Wroc{\l}awski, Kopernika 11, 51-622 Wroc{\l}aw, Poland}
\title{Pulsations in close binaries from the BRITE point of view}
\begin{document}
\maketitle
\begin{abstract}
Using BRITE photometric data for several close binary system we address the problem of damping pulsations in close binary systems due to proximity effects. Because of small number statistics, no firm conclusion is given, but we find pulsations in three relatively close binaries. The pulsations in these binaries have, however, very low amplitudes.
\end{abstract}

\section{Introduction}
One of the main problems in seismic modeling is the lack of precise stellar parameters of the studied objects. For this kind of modeling, masses and radii are the parameters of primary importance. These parameters can be derived when a pulsating star resides in a binary system and the system is both eclipsing and spectroscopically double-lined (SB2). Much less frequent combinations of SB2 and visual orbits also allow deriving masses of the components. This is the case of some luminous and bright stars that have visual interferometric orbits \citep[see, e.g.,][]{2005MNRAS.356.1362D,2009MNRAS.396..842T}. When a binary is wide, the components can be assumed to evolve independently, but such systems usually do not provide the best parameters because orbital periods are long, eclipses are less probable and radial-velocity variations are smaller. In close binary systems, the components usually cannot be considered as evolving independently because tidal effects lead to synchronization of their rotation and circularization of the orbit. In addition, proximity effects cause loss of axial symmetry of the components which itself can have an impact on their structure and evolution. Nevertheless, close binarity usually offers more constraints on stellar parameters. With some credible assumptions (synchronization, rotation axes perpendicular to the orbital plane) equatorial velocities can be derived. Modeling the light and radial velocity curves provides estimates of effective temperatures.

Going back to pulsating stars or rather pulsating components of eclipsing binaries, one can ask how else studies of pulsating stars can benefit from binarity. An interesting possibility is to use eclipse mapping for mode identification \citep[see, e.g.,][]{2011MNRAS.416.1601B}, though this method is difficult to apply, ambiguous in many cases and requires good-quality observations. Binarity is not always an asset; sometimes it poses serious problems for studying pulsating components. First, light and radial-velocity variations due to binarity and pulsations have to be separated. Next, the intrinsic photometric variability of components due to pulsations is diluted by the presence of one or more companions and therefore the observed amplitude is reduced. Finally, one has to identify which component pulsates, which might be difficult if the components are similar.

\section{Do proximity effects damp pulsations?}\label{damp}
In this context, a very interesting problem addressed in this paper arises: what is the impact of close binarity on pulsations? It seems quite obvious that proximity effects should somehow affect pulsations. The reflection effect leads to heating the facing side of a pulsating companion and modifies the structure of its atmosphere. Even more importantly, ellipsoidal effects distort components, which lose their axial symmetry. This problem can be considered for any kind of pulsating stars which are components of eclipsing binaries\footnote{Pulsating stars of practically all known types can reside in binaries, see, e.g, the review by \cite{2006ASPC..349..137P}.}, but we focus here on acoustic ($p$) modes in massive main-sequence stars, i.e.~on $\beta$~Cep-type variability.

The problem whether distortion due to binarity damps pulsations is almost unstudied theoretically. Only recently have \cite{2013MNRAS.434.1869S} published an excellent study of such interaction for a 1~M$_\odot$ model. Therefore, their results may not be fully applicable to massive main-sequence pulsators, but a similar study for more massive stars does not exist. \cite{2013MNRAS.434.1869S} studied how acoustic waves propagate in distorted components of a binary system. They found that the presence of a distortion has a dissipative effect, the larger a star is and the closer it is to filling its Roche lobe. This would indicate that distortion has a damping effect. In addition, they found that some acoustic waves could survive the presence of distortion. In stars nearly filling their Roche lobes, acoustic waves form a ring with an opening angle of about 130$^{\rm o}$ with respect to the inner Langrangian point. As a consequence, the observed amplitude of pulsations will depend on orbital phase. If they exist in real stars, this could be easy to identify observationally by the presence of orbital sidelobes in the frequency spectrum.

For $\beta$~Cep-type stars, the problem was studied observationally by \cite{1983A&A...121...45W}, hereafter WR83. They summarized results of searches for $\beta$~Cep-type pulsations in 17 close binaries with early B-type primaries. These authors, found no pulsation in eleven stars with orbital periods shorter than four days, that is, the orbital period of Spica, at that time the binary with the shortest orbital period and known $\beta$~Cep-type component \citep{1969MNRAS.145..131S}. Therefore, they concluded that there is observational evidence that tidal interaction has a damping effect on $\beta$~Cep-type pulsations.

\section{$\beta$~Cep-type stars in eclipsing binaries}\label{bcec}
During the over 30 years that passed after WR83's study, many other close binaries with $\beta$~Cep components were discovered, and it turned out that the 4-day limit for non-pulsating stars in binaries is no longer compulsory. First, \cite{2001A&A...377..104T} and \cite{2002A&A...394..603S} found pulsations in line profiles of the eclipsing 2.53-day binary $\psi^2$~Ori and non-eclipsing 2.63-day binary $\nu$~Cen, respectively. Then, four $\beta$~Cep stars in eclipsing binaries were found in ASAS photometry \citep{2008A&A...477..917P,2010AN....331.1077D}, including V916~Cen with the orbital period of only 1.46~d, i.e., much shorter than 4 days.

Orbital period itself is not a good measure of stellar distortion. Instead, we propose to use a combination of two parameters, mass ratio, $q = M_{\rm sec}/M_{\rm pr}$, where $M_{\rm pr}$ and $M_{\rm sec}$ are the primary's and secondary's mass, respectively, and fractional radius, $k = R_{\rm pr}/a$, where $R_{\rm pr}$ is the primary's radius and $a$, the separation of the components (or semimajor axis of the relative orbit in the case of an eccentric orbit). According to \cite{1993ApJ...419..344M} and \cite{2007ApJ...670.1326Z}, fractional changes of the total flux of a binary system due to ellipsoidal effect are proportional to $D = qk^3$. We will use $D$ as a measure of distortion of a pulsating primary.

\section{Bright binaries observed by BRITE}
BRITE-Constellation \citep[herafter B-C,][]{2014PASP..126..573W,2016arXiv160800282P}, a fleet of five BRITE nanosatellites, each equipped with a blue or red filter, have observed by now over 300 stars brighter than $V \sim$~5~mag. Among them, there are about two dozen of close binaries with early B-type components, including 13 stars from the WR83 sample. Since BRITE photometry enables one to search for periodic signals with amplitudes down to the sub-mmag detection threshold, a new possibility of verifying the hypothesis of damping effects in close binaries opens up. We present here some preliminary results of the search for pulsations in close binaries based on BRITE data, focusing on one spectacular example, $\delta$~Pictoris.

\subsection{$\delta$~Pictoris}
HD\,42933 = $\delta$~Pic (B0.5 IV, $m_V=$ 4.8 mag) is a bright eclipsing binary discovered as a spectroscopic variable over a hundred years ago \citep{1915LicOB...8..124W}. \cite{1951Obs....71..199C} discovered its photometric variability; subsequent observations allowed a comprehensive spectroscopic \citep{1966MNRAS.131..435T} and photometric \citep{1966MNRAS.131..443C} analysis and derive the system's parameters. The orbital period of the binary amounts to 1.67254~d, while the masses of the components were estimated to be 16.3 and 8.6~M$_\odot$ \citep{1966MNRAS.131..443C}.

Some scarce photometric observations of $\delta$~Pic were made in the following years \citep{1971MNSSA..30...12C,1973MNSSA..32..115K}. The star was also observed by the Hipparcos and ANS satellites. The latter observations were analyzed by \cite{1983PASP...95..319E}. They suggested that in addition to a typical eclipsing light curve, the star shows some additional scatter likely due to intrinsic variability of one of the components. They did not investigate the character of this variability in detail, however. Earlier, \cite{1976IAUS...73..213W} attributed the extra scatter to interstellar matter orbiting in the system.
\begin{figure}[!ht]
\centering
\includegraphics[width=0.75\textwidth]{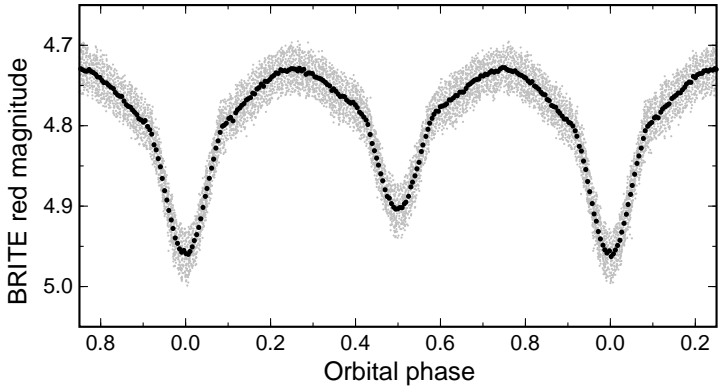}
\caption{Red-filter BRITE light curve of $\delta$~Pic phased with the orbital period of 1.67254~d. The gray points are the individual observations, black dots, averages in adjacent intervals of 0.005 in orbital phase.}
\label{Fig1}
\end{figure}

\begin{figure}[!ht]
\centering
\includegraphics[width=0.75\textwidth]{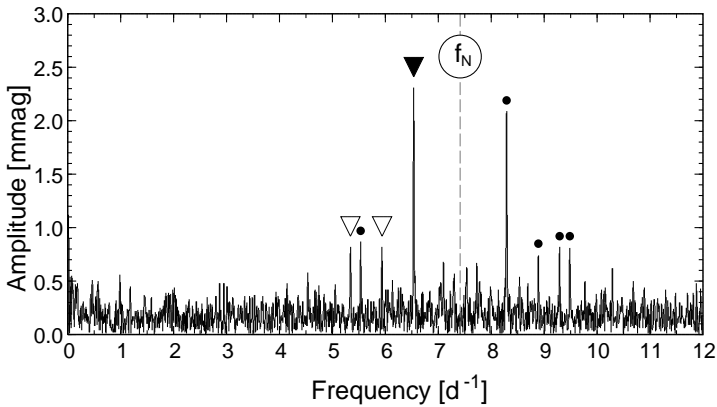}
\caption{Frequency spectrum of BRITE data of $\delta$~Pic after removing variability due to eclipses. The inverted black triangle marks the pulsation frequency $f_1 =$ 6.54~d$^{-1}$, inverted open triangles, two orbital sidelobes, $f_1 - f_{\rm orb}$ and $f_1 - 2f_{\rm orb}$. Dots mark aliases, mostly frequencies mirrored with respect to the orbital Nyquist frequency of BHr, $f_{\rm N}=$ 7.41~d$^{-1}$.}
\label{Fig2}
\end{figure}

BRITE observations of $\delta$~Pic were made by a single BRITE satellite, Heweliusz (BHr), equipped with a red filer. The observations were taken in the $\beta$~Pic field between March 16 and June 2, 2015, resulting in almost 47\,000 individual data points. The eclipsing light curve of $\delta$~Pic is shown in Fig.~\ref{Fig1}. The residuals from the eclipsing light curve were subjected to a time-series analysis. A Fourier frequency spectrum of these residuals is shown in Fig.~\ref{Fig2}. As one can see, the spectrum shows a strong peak at $f_1 \approx$ 6.54~d$^{-1}$ with amplitude of about 2.4~mmag. Given the spectral types of the components, it is very likely a pulsation in a $p$ mode. In addition, two orbital sidelobes can be seen. In the residuals, three more orbital sidelobes were detected. This makes $\delta$~Pictoris a pulsating $\beta$~Cep-type star in a very close 1.67-day binary system. Analysis of the phases and amplitudes of the sidelobes indicate that it is in the primary star where the pulsations originate.
 
\subsection{Other stars}
$\delta$~Pic is not the only close binary in which pulsations were discovered from BRITE data. A detailed analysis of these results is out of the scope of this paper, but we indicate here some preliminary results for five other stars:

{\bf HD\,35411 = \boldmath{$\eta$}~Ori} (B1\,V + B2, $m_V=$ 3.4). The star is a known hierarchical quintuple (or sextuple) consisting of B-type stars, including at least two massive ($\sim$10~M$_\odot$) ones (components Aa, Ab, and possibly Ac, although the last can be itself a binary). The closest pair is eclipsing, with an orbital period of about 8 days. All five BRITEs gathered data for this star, resulting in about 160\,000 data points. The light curves can be fully described by a superposition of an eclipsing light curve and nearly sinusoidal changes due to the proximity effects in component Ac or B. No pulsations exceeding the detection threshold of about 0.3~mmag were found.

{\bf HD\,35715 = \boldmath{$\psi^2$}~Ori} (B1\,III + B2\,V, $m_V=$ 4.6). This is a known ellipsoidal variable \citep{1969JRASC..63..233P} and SB2 binary with an orbital period of 2.526~d in which \cite{2006Ap&SS.304...43S} found a shallow eclipse and fitted a model which predicted also a shallow secondary eclipse. BRITE photometry (all five satellites) confirms the shallow primary eclipse, but no secondary eclipse was found, contrary to the above-mentioned model prediction. This suggests an eccentric orbit. In addition, four significant periodic terms with frequencies in the 8.9\,--\,12.1~d$^{-1}$ range were found. These are likely $p$ modes excited in either component.

{\bf HD\,65818 = V Pup} (B1\,V + B2:, $m_V=$ 4.4). This is a very well known eclipsing binary with a period of 1.454~d and masses of the components equal to 12.5 and 6.1 M$_\odot$ \citep{1998Obs...118..356S}. From the analysis of the O$-$C diagram for the binary period it was suggested that there is an invisible tertiary, possibly even a black hole \citep{2008ApJ...687..466Q}. In BRITE data gathered by all five nanosatellites we detect a single pulsation mode at frequency of 8.8~d$^{-1}$. Similarly to $\delta$~Pic, orbital sidelobes are also detected.

{\bf HD\,120307 = \boldmath{$\nu$}~Cen} (B1\,IV, $m_V=$ 3.4). This is a single-lined spectroscopic binary with a period of 2.625~d \citep{1915LicOB...8..130W}. Non-radial pulsations ($p$ modes) were detected in this star by means of the line profile variability \citep{2002A&A...394..603S}. BRITE data from four satellites reveal no $p$ modes with amplitudes exceeding the detection threshold of about 0.5~mmag.

{\bf HD\,143018 = \boldmath{$\pi$}~Sco} (B1\,V + B2\,V, $m_V=$ 2.8). This is a known occulting double and an SB2 system with orbital period of about 1.57~d \citep{1996Obs...116..387S}. Photometric variability due to proximity effects with the same period is also known. The star is non-eclipsing. BRITE data confirm the variability, but reveal also the presence of a single pulsation $p$ mode and two associated orbital sidelobes.

\section{Conclusions}
\begin{figure}[!ht]
\centering
\includegraphics[width=0.75\textwidth]{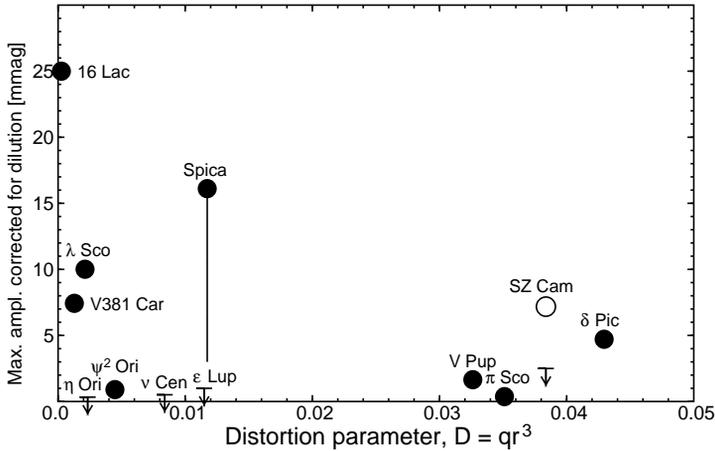}
\caption{Amplitude of the strongest $p$ mode corrected for dilution effect plotted against distortion parameter $D$ (Sect.~\ref{bcec}). Arrows with a horizontal line mark upper limits of non-detections. For SZ Cam, the circle corresponds to the amplitude of the periodic term announced by \cite{2012A&A...539A.139T}; the upper limit plotted below the open circle was estimated from our data \citep{2009AcA....59..349M}.}
\label{Fig3}
\end{figure}

Using results obtained from the analysis of BRITE data and those taken from the literature, we plotted the observed visual semi-amplitudes (or upper detection limits, if $p$ modes were not detected) as a function of dimensionless distortion parameter $D$ introduced in Sect.~\ref{bcec}. The relation is shown in Fig.~\ref{Fig3}. The observed amplitudes were corrected for dilution effects. If distortion indeed damps pulsations in $p$ modes, we should not observe $p$ modes in stars with high $D$ or the amplitudes of pulsations in such stars should be much smaller than for those with smaller $D$. While there is some indication of this kind of behaviour, the picture suffers from small number statistics: the plot is not well populated and no general conclusions can be drawn yet. 

Some points in Fig.~\ref{Fig3} should be explained. The amplitude of the $p$ mode in Spica ($\alpha$~Vir) has decreased over time \citep[][and references therein]{1972MNRAS.156..165S,2016MNRAS.458.1964T}; the filled circle corresponds to the amplitude derived by \cite{1969MNRAS.145..131S}. For SZ~Cam, \cite{2012A&A...539A.139T} announced the discovery of a pulsation with a period of $P=$ 0.33265~d. Since we did our own observations of SZ~Cam \citep{2009AcA....59..349M}, we checked if this variation can be seen in our data. We did not detect this periodicity despite the detection threshold in our data being three times smaller than the amplitude derived by \cite{2012A&A...539A.139T}. In his recent study of SZ~Cam, \cite{2015AstL...41..276G} did not detect this periodicity either. Since the frequency of this term, $f = P^{-1} =$ 3.006~d$^{-1}$ is almost exactly equal to 3~(sidereal days)$^{-1}$, we conclude that this term is almost certainly of instrumental origin.

The present study shows that the question about the damping effect of distortions can be addressed observationally with the use of BRITE data. There are many more eclipsing binaries with early B-type components which have already been or will be observed by B-C. The prospects for improving the statistics in Fig.~\ref{Fig3} look therefore promising. It is interesting to note that all three stars with the highest $D$ in which pulsations in $p$ modes were detected ($\delta$~Pic, $\psi^2$~Ori, $\pi$~Sco) show orbital sidelobes. It will be a matter of detailed modeling to answer the question whether all these sidelobes occur as a result of light dilution and cover a part of a pulsating star during an eclipse or phase-dependent amplitude changes, mentioned in Sect.~\ref{damp}.

\acknowledgements{MR and AP acknowledge support provided by the National Science Center through grants 2015/16/S/ST9/00461 and 2016/21/B/ST9/01126, respectively. This paper is based on data collected by the BRITE Constellation satellite mission, designed, built, launched, operated and supported by the Austrian Research Promotion Agency (FFG), the University of Vienna, the Technical University of Graz, the Canadian Space Agency (CSA), the University of Toronto Institute for Aerospace Studies (UTIAS), the Foundation for Polish Science \& Technology (FNiTP MNiSW), and the National Science Centre (NCN).}


\begin{thebibliography}{34}
\providecommand{\natexlab}[1]{#1}
\providecommand{\url}[1]{\texttt{#1}}
\providecommand{\urlprefix}{URL }
\providecommand{\eprint}[2][]{\url{#2}}

\bibitem[{{B{\'{\i}}r{\'o}} \& {Nuspl}(2011)}]{2011MNRAS.416.1601B}
{B{\'{\i}}r{\'o}}, I.~B., {Nuspl}, J., \emph{{Photometric mode identification
  methods of non-radial pulsations in eclipsing binaries - I. Dynamic eclipse
  mapping}}, \emph{\mnras} \textbf{416}, 1601 (2011)

\bibitem[{{Cousins}(1951)}]{1951Obs....71..199C}
{Cousins}, A.~W.~J., \emph{{Bright variable stars in southern hemisphere (first
  list)}}, \emph{The Observatory} \textbf{71}, 199 (1951)

\bibitem[{{Cousins}(1966)}]{1966MNRAS.131..443C}
{Cousins}, A.~W.~J., \emph{{The eclipsing variable {$\delta$} Pictoris}},
  \emph{\mnras} \textbf{131}, 443 (1966)

\bibitem[{{Cousins} \& {Lagerwey}(1971)}]{1971MNSSA..30...12C}
{Cousins}, A.~W.~J., {Lagerwey}, H.~C., \emph{{UBV observations of variable
  stars}}, \emph{Monthly Notes of the Astronomical Society of South Africa}
  \textbf{30}, 12 (1971)

\bibitem[{{Davis} et~al.(2005)}]{2005MNRAS.356.1362D}
{Davis}, J., et~al., \emph{{Orbital parameters, masses and distance to
  {$\beta$} Centauri determined with the Sydney University Stellar
  Interferometer and high-resolution spectroscopy}}, \emph{\mnras}
  \textbf{356}, 1362 (2005)

\bibitem[{{Drobek} et~al.(2010){Drobek}, {Pigulski}, {Shobbrook}, \&
  {Narwid}}]{2010AN....331.1077D}
{Drobek}, D., {Pigulski}, A., {Shobbrook}, R.~R., {Narwid}, A.,
  \emph{{Photometric study of two {$\beta$} Cephei pulsators in eclipsing
  systems}}, \emph{Astronomische Nachrichten} \textbf{331}, 1077 (2010)

\bibitem[{{Eaton} \& {Wu}(1983)}]{1983PASP...95..319E}
{Eaton}, J.~A., {Wu}, C.-C., \emph{{ANS spectrophotometry -- $\delta$ Pictoris
  as an upper-main-sequence algol system}}, \emph{\pasp} \textbf{95}, 319
  (1983)

\bibitem[{{Gorda}(2015)}]{2015AstL...41..276G}
{Gorda}, S.~Y., \emph{{Results of a long-term monitoring of the multiple system
  SZ Cam}}, \emph{Astronomy Letters} \textbf{41}, 276 (2015)

\bibitem[{{Knipe}(1973)}]{1973MNSSA..32..115K}
{Knipe}, G.~F.~G., \emph{{The eclipsing binary $\delta$ Pictoris}},
  \emph{Monthly Notes of the Astronomical Society of South Africa} \textbf{32},
  115 (1973)

\bibitem[{{Michalska} et~al.(2009){Michalska}, {Pigulski}, {St\k{e}\'slicki},
  \& {Narwid}}]{2009AcA....59..349M}
{Michalska}, G., {Pigulski}, A., {St\k{e}\'slicki}, M., {Narwid}, A., \emph{{A
  CCD search for variable stars of spectral type B in the northern hemisphere
  open clusters. VII. NGC 1502}}, \emph{\actaa} \textbf{59}, 349 (2009),
  \eprint{0910.3672}

\bibitem[{{Morris} \& {Naftilan}(1993)}]{1993ApJ...419..344M}
{Morris}, S.~L., {Naftilan}, S.~A., \emph{{The equations of ellipsoidal star
  variability applied to HR 8427}}, \emph{\apj} \textbf{419}, 344 (1993)

\bibitem[{{Pablo} et~al.(2016)}]{2016arXiv160800282P}
{Pablo}, H., et~al., \emph{{The BRITE Constellation nanosatellite mission:
  Testing, commissioning and operations}}, \emph{ArXiv e-prints}  (2016),
  \eprint{1608.00282}

\bibitem[{{Percy}(1969)}]{1969JRASC..63..233P}
{Percy}, J.~R., \emph{{Light variations in {$\psi$} Orionis}}, \emph{\jrasc}
  \textbf{63}, 233 (1969)

\bibitem[{{Pigulski}(2006)}]{2006ASPC..349..137P}
{Pigulski}, A., \emph{{Intrinsic variability in multiple systems and clusters:
  an overview}}, in C.~{Aerts}, C.~{Sterken} (eds.) Astrophysics of Variable
  Stars, \emph{Astronomical Society of the Pacific Conference Series}, volume
  349, 137 (2006)

\bibitem[{{Pigulski} \& {Poj\-ma{\'n}\-ski}(2008)}]{2008A&A...477..917P}
{Pigulski}, A., {Pojma{\'n}ski}, G., \emph{{{$\beta$} Cephei stars in the
  ASAS-3 data. II. 103 new {$\beta$} Cephei stars and a discussion of
  low-frequency modes}}, \emph{\aap} \textbf{477}, 917 (2008)

\bibitem[{{Qian} et~al.(2008){Qian}, {Liao}, \& {Fern{\'a}ndez
  Laj{\'u}s}}]{2008ApJ...687..466Q}
{Qian}, S.-B., {Liao}, W.-P., {Fern{\'a}ndez Laj{\'u}s}, E., \emph{{Evidence of
  a massive black hole companion in the massive eclipsing binary V Puppis}},
  \emph{\apj} \textbf{687}, 466-470 (2008), \eprint{0806.4944}

\bibitem[{{Schrijvers} \& {Telting}(2002)}]{2002A&A...394..603S}
{Schrijvers}, C., {Telting}, J.~H., \emph{{Identification of non-radial
  pulsation modes in the close-binary $\beta$ Cephei star $\nu$ Centauri}},
  \emph{\aap} \textbf{394}, 603 (2002)

\bibitem[{{Shobbrook} et~al.(1969){Shobbrook}, {Herbison-Evans}, {Johnston}, \&
  {Lomb}}]{1969MNRAS.145..131S}
{Shobbrook}, R.~R., {Herbison-Evans}, D., {Johnston}, I.~D., {Lomb}, N.~R.,
  \emph{{Light variations in Spica}}, \emph{\mnras} \textbf{145}, 131 (1969)

\bibitem[{{Shobbrook} et~al.(1972){Shobbrook}, {Lomb}, \&
  {Herbison-Evans}}]{1972MNRAS.156..165S}
{Shobbrook}, R.~R., {Lomb}, N.~R., {Herbison-Evans}, D., \emph{{The short
  period light and velocity variations in $\alpha$ Virginis.}}, \emph{\mnras}
  \textbf{156}, 165 (1972)

\bibitem[{{Shobbrook} \& {Zola}(2006)}]{2006Ap&SS.304...43S}
{Shobbrook}, R.~R., {Zola}, S., \emph{{Photometric studies of bright southern
  binary systems: {$\epsilon$} Cra and {$\psi$} Ori}}, \emph{\apss}
  \textbf{304}, 43 (2006)

\bibitem[{{Springer} \& {Shaviv}(2013)}]{2013MNRAS.434.1869S}
{Springer}, O.~M., {Shaviv}, N.~J., \emph{{Asteroseismic effects in close
  binary stars}}, \emph{\mnras} \textbf{434}, 1869 (2013)

\bibitem[{{Stickland} et~al.(1996){Stickland}, {Lloyd}, {Koch}, \&
  {Pachoulakis}}]{1996Obs...116..387S}
{Stickland}, D.~J., {Lloyd}, C., {Koch}, R.~H., {Pachoulakis}, I.,
  \emph{{Spectroscopic binary orbits from ultraviolet radial velocities Paper
  23: $\pi$ Scorpii (HD 143018)}}, \emph{The Observatory} \textbf{116}, 387
  (1996)

\bibitem[{{Stickland} et~al.(1998){Stickland}, {Lloyd}, {Pachoulakis}, \&
  {Koch}}]{1998Obs...118..356S}
{Stickland}, D.~J., {Lloyd}, C., {Pachoulakis}, I., {Koch}, R.~H.,
  \emph{{Spectroscopic binary orbits from ultraviolet radial velocities. Paper
  29: V Puppis (HD 65818)}}, \emph{The Observatory} \textbf{118}, 356 (1998)

\bibitem[{{Tamajo} et~al.(2012)}]{2012A&A...539A.139T}
{Tamajo}, E., et~al., \emph{{Asiago eclipsing binaries program IV. SZ
  Camelopardalis, a {$\beta$} Cephei pulsator in a quadruple, eclipsing
  system}}, \emph{\aap} \textbf{539}, A139 (2012), \eprint{1202.0130}

\bibitem[{{Tango} et~al.(2009)}]{2009MNRAS.396..842T}
{Tango}, W.~J., et~al., \emph{{A new determination of the orbit and masses of
  the Be binary system {$\delta$} Scorpii}}, \emph{\mnras} \textbf{396}, 842
  (2009)

\bibitem[{{Telting} et~al.(2001){Telting}, {Abbott}, \&
  {Schrijvers}}]{2001A&A...377..104T}
{Telting}, J.~H., {Abbott}, J.~B., {Schrijvers}, C., \emph{{Apsidal motion and
  non-radial pulsations in $\psi^2$ Ori}}, \emph{\aap} \textbf{377}, 104 (2001)

\bibitem[{{Thackeray}(1966)}]{1966MNRAS.131..435T}
{Thackeray}, A.~D., \emph{{The spectroscopic binary {$\delta$} Pictoris}},
  \emph{\mnras} \textbf{131}, 435 (1966)

\bibitem[{{Tkachenko} et~al.(2016)}]{2016MNRAS.458.1964T}
{Tkachenko}, A., et~al., \emph{{Stellar modelling of Spica, a high-mass
  spectroscopic binary with a {$\beta$}~Cep variable primary component}},
  \emph{\mnras} \textbf{458}, 1964 (2016)

\bibitem[{{Waelkens} \& {Ru\-fe\-ner}(1983)}]{1983A&A...121...45W}
{Waelkens}, C., {Rufener}, F., \emph{{An observational study of the influence
  of close companions on the pulsations of $\beta$ Cephei stars}}, \emph{\aap}
  \textbf{121}, 45 (1983)

\bibitem[{{Weiss} et~al.(2014)}]{2014PASP..126..573W}
{Weiss}, W.~W., et~al., \emph{{BRITE-Constellation: Nanosatellites for
  Precision Photometry of Bright Stars}}, \emph{\pasp} \textbf{126}, 573 (2014)

\bibitem[{{Wilson}(1915{\natexlab{a}})}]{1915LicOB...8..124W}
{Wilson}, R.~E., \emph{{Fourteen stars whose radial velocities vary}},
  \emph{Lick Observatory Bulletin} \textbf{8}, 124 (1915{\natexlab{a}})

\bibitem[{{Wilson}(1915{\natexlab{b}})}]{1915LicOB...8..130W}
{Wilson}, R.~E., \emph{{The orbit of the spectroscopic binary $\nu$ Centauri}},
  \emph{Lick Observatory Bulletin} \textbf{8}, 130 (1915{\natexlab{b}})

\bibitem[{{Wu}(1976)}]{1976IAUS...73..213W}
{Wu}, C.-C., \emph{{Ultraviolet observations of some close binary systems by
  the Astronomical Netheriands Satellite -- ANS}}, in P.~{Eggleton},
  S.~{Mitton}, J.~{Whelan} (eds.) Structure and Evolution of Close Binary
  Systems, \emph{IAU Symposium}, volume~73, 213 (1976)

\bibitem[{{Zucker} et~al.(2007){Zucker}, {Mazeh}, \&
  {Alexander}}]{2007ApJ...670.1326Z}
{Zucker}, S., {Mazeh}, T., {Alexander}, T., \emph{{Beaming binaries: a new
  observational category of photometric binary stars}}, \emph{\apj}
  \textbf{670}, 1326 (2007)

\end{thebibliography}
\end{document}